%% file: main.tex
\documentclass[sigconf]{acmart}

\input{packages.tex}
\input{cmds.tex}

\input{shorthand.tex}
\usepackage{hyperref}
\usepackage[T1]{fontenc}
\usepackage{etoolbox}
\usepackage{url}            
\titlespacing*{\section}
{0pt}{.0ex}{0.0ex}
\titlespacing*{\subsection}
{0pt}{0.0ex}{0.0ex }
\def\BibTeX{{\rm B\kern-.05em{\sc i\kern-.025em b}\kern-.08emT\kern-.1667em\lower.7ex\hbox{E}\kern-.125emX}}

\acmYear{2019}\acmVolume{37}\acmNumber{4}\acmArticle{111}\acmMonth{8}
\acmISBN{978-1-4503-5720-3}
\acmDOI{10.1145/1122445.1122456}
\title{Reinforcement Learning based Interconnection Routing for Adaptive Traffic Optimization} 
\author{Sheng-Chun Kao\textsuperscript{*1}, Chao-Han Huck Yang\textsuperscript{*1}, Pin-Yu Chen\textsuperscript{2}, Xiaoli Ma\textsuperscript{1}, Tushar Krishna\textsuperscript{1}}\thanks{\textsuperscript{*}Equal Contribution. Code: \url{github.com/huckiyang/interconnect-routing-gym}}
\affiliation{\institution{\textsuperscript{1}Georgia Institute of Technology\hspace{20pt} \textsuperscript{2}IBM Watson AI Foundation Group}}
\email{{skao6,   huckiyang}@gatech.edu, pin-yu.chen@ibm.com, {xiaoli,tushar}@ece.gatech.edu}

\setcopyright{rightsretained} 
\begin{document}
\renewcommand{\shortauthors}{Sheng-Chun Kao, et al.}
\input {\SECDIR/00-Abstract.tex}
\copyrightyear{2019} 
\acmYear{2019} 
\acmConference[NOCS '19]{International Symposium on Networks-on-Chip}{October 17--18, 2019}{New York, NY, USA}
\acmBooktitle{International Symposium on Networks-on-Chip (NOCS '19), October 17--18, 2019, New York, NY, USA}
\acmDOI{10.1145/3313231.3352369}
\acmISBN{978-1-4503-6700-4/19/10}
\maketitle
\thispagestyle{firstpage}
\pagestyle{plain}

%
\input {\SECDIR/01-Introduction.tex}
\input {\SECDIR/03-Algorithm.tex}

\input {\SECDIR/04-Modules.tex}
\input {\SECDIR/05-Experiment.tex}
\input {\SECDIR/07-Conclusion.tex}

\bibliographystyle{unsrt}
\bibliographystyle{ACM-Reference-Format}
\bibliography{main}

\end{document}

%% file: packages.tex
\usepackage{epsfig}
\usepackage{graphicx}
\usepackage{amsmath}
\usepackage{amssymb}
\usepackage{helvet}  
\usepackage{courier}  
\usepackage{url}  
\usepackage{xcolor}
\usepackage{multirow}
\usepackage{subcaption}
\usepackage{array}
\usepackage{algorithm}
\usepackage{algorithmic}
\usepackage{fullpage}
\usepackage{times}
\usepackage{xcolor}
\usepackage{listings}
\usepackage{titlesec}

%% file: cmds.tex
\newcommand{\PCignore}[1]{}

\newcommand{\squishlist}{
 \begin{list}{$\bullet$}
  { \setlength{\itemsep}{0pt}
     \setlength{\parsep}{3pt}
     \setlength{\topsep}{3pt}
     \setlength{\partopsep}{0pt}
     \setlength{\leftmargin}{1.5em}
     \setlength{\labelwidth}{1em}
     \setlength{\labelsep}{0.5em} } }

\newcommand{\squishlisttwo}{
 \begin{list}{$\bullet$}
  { \setlength{\itemsep}{0pt}
     \setlength{\parsep}{0pt}
    \setlength{\topsep}{0pt}
    \setlength{\partopsep}{0pt}
    \setlength{\leftmargin}{2em}
    \setlength{\labelwidth}{1.5em}
    \setlength{\labelsep}{0.5em} } }

\newcommand{\squishend}{
  \end{list}  }

%% file: sections/00-Abstract.tex
\begin{abstract}
Applying Machine Learning (ML) techniques to design and optimize computer architectures 
is a promising research direction.
Optimizing the runtime
performance of a Network-on-Chip (NoC) necessitates 
a continuous learning framework.
In this work, we demonstrate the promise of 
applying reinforcement learning (RL) to optimize NoC runtime performance.
We present three RL-based methods for learning optimal routing algorithms. The experimental results show the algorithms can successfully learn a near-optimal solution across different environment states.
\end{abstract}

\keywords{Network-on-Chips, Intelligent Physical Systems, Reinforcement Learning, Congestion Control, Scalable Modeling.}

%% file: sections/01-Introduction.tex
\begin{figure}[ht]
\begin{center}
\includegraphics[width=1.0\linewidth]{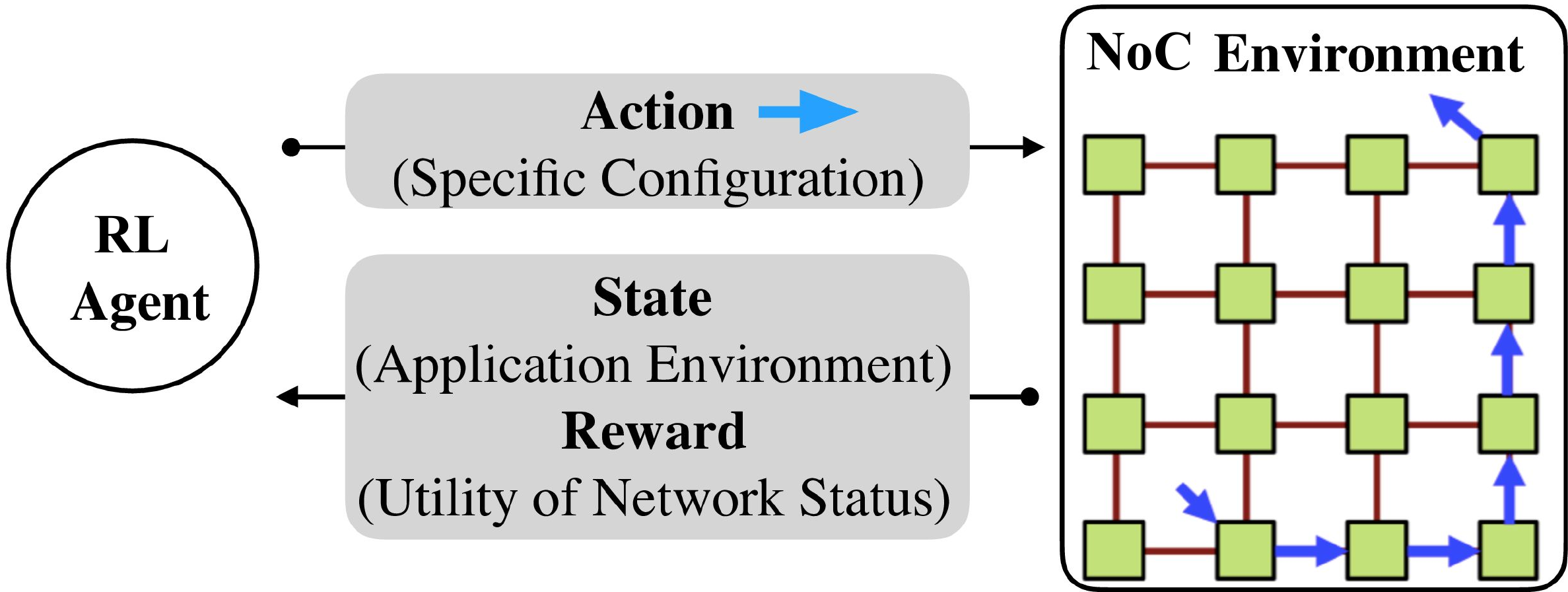}
\end{center}
   \caption{Proposed NoC reinforcement learning scheme}
\vspace{-0.08cm}
\label{fig:figure1}
\end{figure}

\section{Introduction}
Researchers have started applying machine learning (ML) algorithms for optimizing the runtime performance of computer systems~\cite{rl_google}.
Networks-on-chip (NoCs) form the 
communication backbone of many-core systems; 
learning traffic behavior and optimizing 
the latency and bandwidth characteristics 
of the NoC in response 
to runtime changes is a 
promising candidate for applying ML.
This work 
explores opportunities that Reinforcement learning (RL) techniques \cite{sutton1998reinforcement} provide for learning optimal routing algorithms for varying traffic within a NoC.

RL techniques work via continuous interactions with an environment to 
learn the optimal policy. They have demonstrated  promising results in robotics ~\cite{brockman2016openai}, playing Atari games, and computer network traffic control \cite{kong2018improving}.  In this work, we study how classical RL algorithms 
work for NoC 
routing and develop a framework for applying these RL algorithms to NoCs. We further present an extended OpenAI Gym package for studying RL-based routing control in NoC simulations based on gem5~\cite{binkert2011gem5}. Our results show the RL agents were able to learn and pick the optimal routing algorithm for a traffic pattern to maximize a customized network objective such as the routing throughput. 


%% file: sections/03-Algorithm.tex
\begin{figure*}[ht]
\begin{center}
\includegraphics[width=1.0\linewidth]{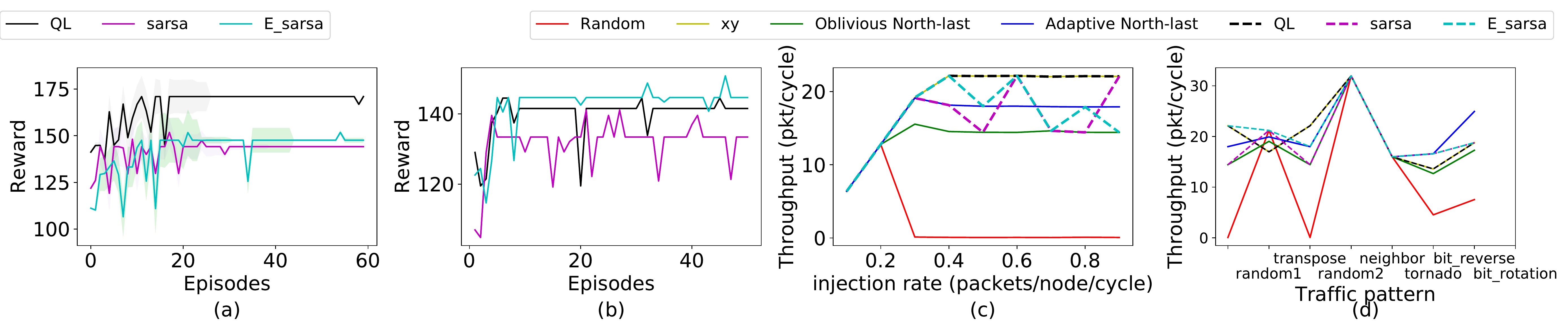}
\end{center}
\vspace{-0.40cm}
   \caption{Reward over training episodes on (a) the $Case 1$ and (b) the $Case 2$ with standard deviation; throughput of fixed routing algorithms and RL-selected routing on (c) the $Case 1$ and (d) the $Case 2$.}
\vspace{-0.2cm}
\label{fig:figCobmine}
\end{figure*}

\section{RL-based Routing Optimization}
\textbf{Overview.} We develop a framework to use RL to optimize NoC routing decision. As shown in Fig. \ref{fig:figure1}, in NoC environment, our RL agent keeps records of the current network state and its corresponding reward (throughput), and then suggests an action (a choice of routing algorithms) with the highest expected reward, based on the learned information. 



\textbf{Target Task.} The goal of our RL agent is to learn an optimal routing algorithm that maximizes throughput for the current application.



\textbf{Defining Utility Function for NoCs.} RL works by optimizing actions for a utility or reward function. 
It treats the problem as a Markov process, which means if we have current state with the learned information, we can decide future action to optimize reward.
%
In our use case for NoCs, we define the utility objective  ($\mathbf{U_{1}}$) by calculating the throughput:  
\begin{equation}
    U_{1} \equiv Throughput =\frac{ Number~of~Packets_{received}}{Number~of~ Cycles_{execution}} .
    \label{eq:eq1}
\end{equation}

\textbf{Proposed RL Framework.}
As a central motivation in RL, value function approaches attempt to find a policy that maximizes the return by maintaining a set of estimates of expected reward.  We use the designed Utility functions in Equation~\ref{eq:eq1}.
The RL agent's (Fig \ref{fig:figure1}) action selection is modeled as a policy ($\mathbf{\pi}$):
\begin{equation}
\begin{array}{l}{\pi : S \times A \rightarrow[0,1]} \\ {\pi(a | s)=\operatorname{Pr}\left(a_{t}=a | s_{t}=s\right).}\end{array}
\end{equation}
where the return $\mathbf{R}$ could be calculated:
\begin{equation}
R=\sum_{t=0}^{\infty} \gamma^{t} U_{t},
\end{equation}
$U_{t}$ is the temporal utility measured at the time $t$, and $\gamma$ is the discount factor in the Markov process. The action-value function of such an optimal policy $Q^{\pi}$ is called the optimal action-value function to attain maximum expectation of $\mathbf{R}$ as:
\begin{equation}
Q^{\pi}(s, a)=E[R | s, a, \pi] .
\end{equation}

\textbf{RL Algorithms.}
We consider three temporal differential approaches: Q-learning, SARSA, and Expected-SARSA. We do not use deep reinforcement learning (DRL) methods owing to a high real-time memory consumption of 
DRL from previous studies \cite{sutton1998reinforcement} which make them prohibitive.

%% file: sections/04-Modules.tex
\section{Experimental Methodology}
\textbf{Extending OpenAI Gym for Interconnection Routing.}\\
OpenAI $Gym$~\cite{brockman2016openai} is a benchmark suite for developing RL algorithms. Consequently, we provide a first scalable environment for fast prototyping new RL-integrated NoCs, called $interconnet$-$routing$-$gym$ ($icr$-$gym$). Our proposed $icr$-$gym$ environment includes:
\begin{itemize}
    \item $State$ -- gem5 statistics with the injected flits, received flits, and average latency
    \item $Action$ -- a set of standard routing algorithms (e.g., xy, oblivious north-last, adaptive-north-last, random-adaptive) for RL agents to choose from
    \item $Reward$ -- a customized network objective(s)  (e.g., latency and throughput) of a selected NoC topology
    \item $Info$ -- Boolean format for thresholding at desired reward    
\end{itemize}
\vspace{-0.12cm}
\textbf{Case Study 1: NoCs with Incremental Injection Rate.}\\
In this scenario, we use Garnet2.0 \cite{agarwal2009garnet}, an NoC simulator network model inside gem5 \cite{binkert2011gem5}. We provide a target topology as an 8-by-8 mesh. We start  packet injection at a low rate and then increase the rate as time goes on. Our goal is to optimize the performance by choosing optimal routing algorithms from the action space at each transition of environment state. We set the action space of $icr$-$gym$ in both two case studies as four choices: random routing, xy routing, oblivious North-last, and  adaptive North-last (which uses the number of free virtual channels at the next router as a proxy for choosing the output port).
\\
\textbf{Case Study 2: NoCs with Dynamic Traffic Patterns.}\\
In this scenario, we simulate the workload of a data center network. 
For example, in a Google data center, the primary application could change from mail service to video traffic in the different time frame of the day. Therefore, we simulate this scenario by switching from one network traffic pattern to another. We use seven different synthetic traffic patterns provided by Garnet2.0 in the experiments, e.g., random, transpose, and bit reversed traffic, as shown in  Fig. \ref{fig:figCobmine} (d). Then our environment is defined as the continuously changing network traffic NoC. 
We apply RL to optimize the routing algorithm decision at each state transition.

%% file: sections/05-Experiment.tex
\section{Evaluation}
In $Case 1$, the reward is defined as throughput. 
The reward feedback in each episode is shown in  Fig. \ref{fig:figCobmine} (a). We can observe that the rewards of the three RL algorithms converge and are stable. We examine our learned models by testing their throughput through one episode of an entire state transition which is $\mathbf{0.1}$ to $\mathbf{0.9}$, as described in $Case 1$. 
We compare the throughput of NoCs guided by our RL agents in different injection rates with the throughput of fixed baseline routing algorithms (e.g., random routing, xy routing), as shown in Fig. \ref{fig:figCobmine} (c). For example, the throughput of random routing saturates to near $\mathbf{0}$ after rate $\mathbf{0.3}$ because of deadlock. Oblivious North-last avoids deadlock and saturates at higher throughput. As for our RL method, the Sarsa chooses Adaptive North-last from rate $\mathbf{0.1}$ to $\mathbf{0.4}$ and oblivious North-last at rate $\mathbf{0.5}$. However, the QL always makes the optimal choice out of four routing algorithms.

In $Case 2$, we have different traffics patterns (e.g., random, tornado traffics) as our states. we show the result under the injection rate of $\mathbf{0.5}$ in Fig. \ref{fig:figCobmine} (b), and the results under different injection rates can all converge and follow the same trend. The throughput of an entire state transition is in  Fig. \ref{fig:figCobmine} (d). 
We could observe all three RL methods deliver near optimal choices across all states. Through theses experiments, we show that our method could serve as a decision agent for the data center facing various workloads.

%% file: sections/07-Conclusion.tex
\section{Conclusion}
We develop and demonstrate 
a framework to apply RL to act as a continually learning agent that configures the routing algorithms decision in a NoC.
We concretely show the effectiveness of policy-based RLs on NoC problems. We hope this work will inspire future extensions to bring more RL algorithms to a wide range of NoC problems for the computer networks community.

%% file: main.bbl
\begin{thebibliography}{1}

\bibitem{rl_google}
Azalia Mirhoseini et~al.
\newblock Device placement optimization with reinforcement learning.
\newblock In {\em Proceedings of the 34th International Conference on Machine
  Learning-Volume 70}, pages 2430--2439, 2017.

\bibitem{sutton1998reinforcement}
Richard~S Sutton et~al.
\newblock {\em Reinforcement learning: An introduction}.
\newblock MIT press, 1998.

\bibitem{brockman2016openai}
Greg Brockman et~al.
\newblock Openai gym.
\newblock {\em arXiv preprint arXiv:1606.01540}, 2016.

\bibitem{kong2018improving}
Yiming Kong et~al.
\newblock Improving tcp congestion control with machine intelligence.
\newblock In {\em ACM NetAI}, pages 60--66, 2018.

\bibitem{binkert2011gem5}
Nathan Binkert et~al.
\newblock The gem5 simulator.
\newblock {\em ACM SIGARCH Computer Architecture News}, 39(2):1--7, 2011.

\bibitem{agarwal2009garnet}
Niket Agarwal et~al.
\newblock Garnet: A detailed on-chip network model inside a full-system
  simulator.
\newblock In {\em 2009 IEEE international symposium on performance analysis of
  systems and software}, pages 33--42. IEEE, 2009.

\end{thebibliography}
